\documentclass[fleqn,10pt]{wlscirep}

\newcommand{\onlinecite}[1]{\hspace{-1 ex} \nocite{#1}\citenum{#1}}
\DeclareMathAlphabet\mathbfcal{OMS}{cmsy}{b}{n}
\title{Bulk band inversion and surface Dirac cones in LaSb and LaBi : Prediction of a new topological heterostructure}

\author[1,*]{Urmimala Dey}
\author[1]{Monodeep Chakraborty}
\author[1,2,3]{A. Taraphder}
\author[4]{Sumanta Tewari}
\affil[1]{Centre for Theoretical Studies, Indian Institute of Technology Kharagpur, Kharagpur-721302, India}
\affil[2]{Department of Physics, Indian Institute of Technology Kharagpur, Kharagpur-721302, India}
\affil[3]{School of Basic Sciences, Indian Institute of Technology Mandi, HP 175005 India}
\affil[4]{Department of Physics and Astronomy, Clemson University, Clemson, South Carolina 29634, USA}

\affil[*]{urmimaladey@iitkgp.ac.in}

\begin{abstract}
We perform \textit{ab initio} investigations of the bulk and surface band structures of LaSb and LaBi and resolve the existing disagreements about the topological property of LaSb, considering LaBi as a reference. We examine the bulk band structure for band inversion, along with the stability of surface Dirac cones (if any) to time-reversal-preserving perturbations, as a strong diagnostic test for determining the topological character of LaSb, LaBi and LaSb-LaBi multilayer. 
A detailed \textit{ab initio} investigation of a multilayer consisting of alternating unit cells of LaSb and LaBi shows the presence of band inversion in the bulk and a massless Dirac cone on the (001) surface, which remains stable under the influence of time-reversal-preserving perturbations, thus confirming the topologically non-trivial nature of the multilayer in which the electronic properties can be tailored as per requirement. A detailed $\mathbb{Z}_2$ invariant calculation is performed to arrive at a holistic conclusion.
\end{abstract}
\begin{document}

\flushbottom
\maketitle
%
%
\thispagestyle{empty}


\section*{Introduction}

Topological insulators (TI) as a new class of insulators with time reversal (TR) symmetry and topologically protected surface states have recently drawn much attention in condensed matter physics and materials science.~\cite{Mele,Moore,Fu:2007,Roy,Bernevig,Konig,Hsieh,Kane_2010, Zhang_2011} TIs are different from conventional (topologically trivial) band insulators due to the presence of a spin-orbit-driven band inversion in the bulk spectrum, whereby the usual ordering of the conduction and valence bands is inverted in the reciprocal space. The presence of TR symmetry in these systems allows the definition of an Ising or $\mathbb{Z}_2$ topological index, which mathematically distinguishes the wave functions of TIs from those of ordinary band insulators. The band inversion in the bulk spectrum and the associated non-trivial $\mathbb{Z}_2$ index ensure that the bulk band structure of three-dimensional (3D) strong TIs is crossed by an odd number of pairs of topologically protected gapless surface states,
which come in the form of an odd number of Dirac cones on a given surface.
Furthermore,
the electron spins in the surface state branches are aligned perpendicular to their momenta contributing an overall Berry's phase of $\pi$ to the fermion wave functions.
In addition to various anomalous quantum phenomena and the associated fundamental scientific interest,~\cite{Peng_2010,Chen_2010,He_2011,Hor_2011,Qu_2010} the topological protection and the non-trivial spin textures of the Dirac-cone-like surface states in TI can be of interest for spintronic and quantum computation applications.~\cite{Kane_2010,Zhang_2011} For other topological systems supporting topologically non-trivial Dirac-cone-like bulk and surface states see Refs.~[\onlinecite{Ando_2013,Bansil_2012,Tanaka_2012,Hasan_2012,Story_2012,Hasan_2014,Gibson_2014,Liu_2014,Hasan_2015,Ma_2015, Yang_2015,Ando_2016,Schoop_2016,Takane_2016}].

Conventional insulators can also have gapless surface states localized at boundaries separating the bulk insulator from vacuum. However, in contrast to TIs, the surface states of these systems are highly sensitive to disorder, leading to Anderson localization and surface reconstruction. While the surface states in 3D strong TIs are topologically protected by the TR symmetry, even time-reversal-preserving perturbations can eliminate the gapless surface states in topologically trivial band insulators. The topological protection of the surface Dirac cones in 3D strong TIs is a direct consequence of the bulk band inversion. In inversion symmetric systems, the $\mathbb{Z}_2$ index of TR-invariant band insulators~\cite{Fu:2007,Fu:Inversion} can be written as the product of the parities of the filled energy bands at the TR-invariant momenta (TRIM). Then, an odd number of inversions between a pair of bands of opposite parities in the bulk lead to a non-trivial $\mathbb{Z}_2$ index, resulting in gapless, topologically protected, Dirac cone-like surface states at the boundary between 3D TI and vacuum, which can be viewed as a trivial insulating medium. 
In contrast, the gapless accidental surface Dirac cones which may exist in 3D conventional band insulators enjoy no such protection, and can in principle be gapped out even by TR-preserving perturbations. Thus, a critical examination of the bulk band structure for band inversion, along with examining the stability of surface Dirac cones (if any) to TR-preserving perturbations, constitute a strong diagnostic test for determining the topological character of 3D TR-invariant insulators. In this paper, we examine the presence of bulk band inversion and stability of surface Dirac cones in LaSb as well as LaBi, a proven $\mathbb{Z}_2$ semimetal, in order to resolve the existing disagreements
 among various groups regarding the topological character of LaSb through  our first-principles calculations. Detailed \textit{ab initio} calculation of the bulk and surface band structures of a multilayer consisting of alternating unit cells of LaSb and LaBi confirm the topological properties of such a multilayer, despite the fact that LaSb itself is topologically trivial. Such artificially engineered topological multilayer heterostructure can be a promising playground for realizing novel topological functionalities and device applications.

The proposal of topological nature of lanthanum monopnictides (LaX : X = N, P, As, Sb, Bi) by Zeng \textit{et al}~\cite{Bansil_2015} ignited enormous interest, which
led to thorough investigations of the topological aspects of rare-earth monopnictides.~\cite{Guo_2016,Guo_2017,Neupane_2016,Felser_2016,Kaminski_2017} Among them, LaSb and LaBi have been
classified as extreme magnetoresistive (XMR) materials with unusual resistivity plateau.~\cite{Gibson_2015,Sun_2016} Both theoretical and experimental studies have confirmed
the topologically non-trivial nature of LaBi supported by the presence of a band inversion in the bulk band structure, though there are some controversies on the number and nature of
 the surface Dirac cones.~\cite{Lou_2017,Nayak_2017}
However, the topological character of LaSb remains highly debatable. The first-principles calculations of the bulk band structure using semi-local functionals, for example, GGA (generalized gradient approximation) or LDA (local density approximation) along with spin-orbit coupling, shows that bulk LaSb is topologically non-trivial due to the presence of band
inversion near the $X$-point.~\cite{Guo_2016} However, the meta-GGA calculation with modified Becke Johnson (mBJ) potential obliterates the bulk band inversion and renders LaSb topologically
trivial.~\cite{Guo_2016} On the other hand, Niu \textit{et al}~\cite{Niu_2016} have found Dirac cones at the (001) surface using VUV-ARPES experiments. However, the  ARPES study by Zeng \textit{et al}~\cite{Zeng_2016} suggests topologically trivial nature of LaSb, raising further apprehensions about the disagreements among various groups regarding the topological character of LaSb. 

Recent bulk sensitive ARPES experiments using
soft-x-ray photons by Oinuma \textit{et al}~\cite{Oinuma_2017} have shown that there is no band inversion present in the bulk band structure of LaSb, leading to the conjecture that the
Dirac-cone-like features on the (001) surface may be topologically trivial. 
However, to the best of our knowledge, till date no slab calculation examining the surface band structure of LaSb exists which can confirm the actual topological character of LaSb surface states. First-principles calculations of the bulk band structure confirming the presence or absence of band inversion, along with examining the robustness of surface Dirac cones, if any, to TR-preserving perturbations, are needed to ascertain the true topological character of LaSb in comparison to LaBi, which is known as a topologically non-trivial $\mathbb{Z}_2$ semimetal.~\cite{Lou_2017,Nayak_2017}

In this work, we first calculate the bulk band structures of LaSb and LaBi with and without the meta-GGA functional to check the robustness of the band inversions.
Although LDA+SO calculations for LaSb show clear evidence of band inversion, adding an mBJ potential to our spin-orbit incorporated LDA calculations for bulk LaSb we no longer observe any signature of bulk band inversion. By contrast, even upon including the mBJ potential in the LDA+SO calculation for LaBi, we find that the bulk band inversion survives and in fact is further consolidated by reducing the gap between the conduction and the valence bands.
We then proceed to analyze the surface states of LaSb with our slab calculations and find an odd number of Dirac cones on the (001) surface. Crucially, to check the stability of the surface Dirac cone, we apply a time-reversal-preserving perturbation in the form of uniaxial strain, which results in a gapped surface state spectrum in LaSb. This leads us to conclude that the unusual Dirac-cone-like surface states previously observed in LaSb by VUV-ARPES experiments~\cite{Niu_2016} are topologically trivial and arise due to accidental degeneracy. 
By contrast, we find that the gapless surface state Dirac spectrum in LaBi remains stable when subjected to the same TR-preserving perturbations. To the best of our knowledge, the stability of the Dirac cones found in the (001) surface spectrum of LaBi against TR-preserving perturbations has not been put to test previously. Comparing and contrasting the bulk as well as slab calculations for LaSb and LaBi, where the latter is a known topological $\mathbb{Z}_2$ semimetal,~\cite{Lou_2017,Nayak_2017} captures the crucial difference between the two systems regarding the topological robustness of band inversion and surface Dirac cones, substantiating our conclusion about the topological character of LaSb. 

After probing the topological properties of LaSb and LaBi individually, including the respective slab calculations, we study a multilayer heterostructure fabricated from   
these two systems, with an objective to tune the topological properties of LaSb. 
To that end, we have constructed a multilayer consisting of alternate unit cells of LaSb and LaBi stacked 
along the (001) direction. Interestingly, the LDA calculation including spin-orbit 
coupling (SOC) shows that the conduction band and the valence band
 get inverted near the $X$-point in the bulk Brillouin zone, which survives the effect of mBJ potential. In order to confirm the topological nature of the multilayer, we also calculate the (001)-projected surface band 
dispersion. An odd number of Dirac cones appears near the $\bar{M}$-point, which remains gapless even under the influence of a TR-preserving perturbation, which confirms the topologically robust nature of the LaSb-LaBi heterostructure. Our prediction of a new topological multilayer heterostructure involving LaSb and LaBi 
holds promise of new device applications where functionalities can be
manipulated by chemical substitution and/or altering the stacking sequence of layers in a multilayer. 

The paper is organized as follows. In section II, we present the computational methods used for the band structure calculations. The bulk and surface spectra of LaSb  are calculated and presented in section III. 
In section IV, we discuss the bulk as well as surface band structures of LaBi, which is a known $\mathbb{Z}_2$ semimetal,~\cite{Lou_2017,Nayak_2017} and compare them with those of LaSb. In contrast to LaSb, we find that the bulk band inversion survives in LaBi even after adding the mBJ potential on the spin-orbit incorporated LDA calculations. Moreover, in LaBi the surface Dirac cone found at the (001)-$\bar{M}$ point survives the effect of a TR-preserving perturbation. This robustness of the surface states against TR-preserving perturbations is a signature of the topologically non-trivial nature of LaBi while we find that LaSb is topologically trivial. In section V, we consider a heterostructure made of alternating unit cells of LaSb and LaBi along the (001) direction. Presence of bulk band inversion, appearance of an odd number of gapless Dirac cones at the (001) surface, and the robustness of the surface Dirac cones to TR-preserving perturbations establish the topologically non-trivial nature of LaSb-LaBi multilayer heterostructure.
\section*{Bulk and surface spectrum of LaSb}
\subsection*{Bulk band structure of LaSb}
\begin{figure}[h]
\centering
\includegraphics[scale=.15]{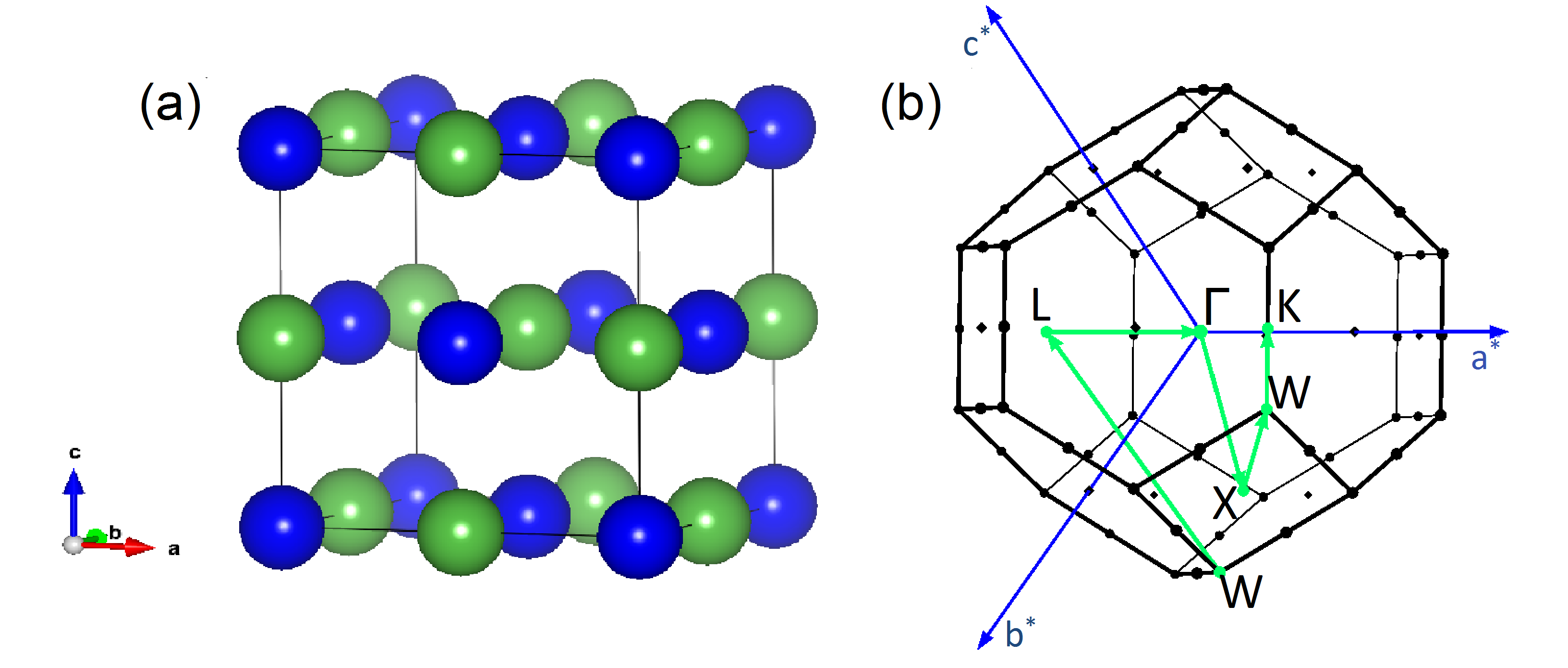}
\caption{(a) The FCC unit cells of LaSb or LaBi. The green spheres denote the La atoms and the blue spheres are the Sb or Bi atoms. (b) FCC bulk Brillouin zone. $a^{*}$, $b^{*}$, $c^{*}$ denote the reciprocal lattice vectors.}
\label{LaBi_struct}
\end{figure}

LaSb is known to crystallize in a face centered cubic structure belonging to the space group $Fm\bar{3}m$ (No. 225) in which the La and Sb atoms occupy
($0$,$0$,$0$) and  ($\frac{1}{2},\frac{1}{2},\frac{1}{2}$) positions respectively. The equilibrium lattice parameter $a$ = 6.402 \r{A}, obtained after volume optimization, is used for the band structure calculations. The crystal structure of LaSb and the bulk Brillouin zone are shown in Fig.~\ref{LaBi_struct}.

\begin{figure}[h]
\centering
\includegraphics[scale=.25]{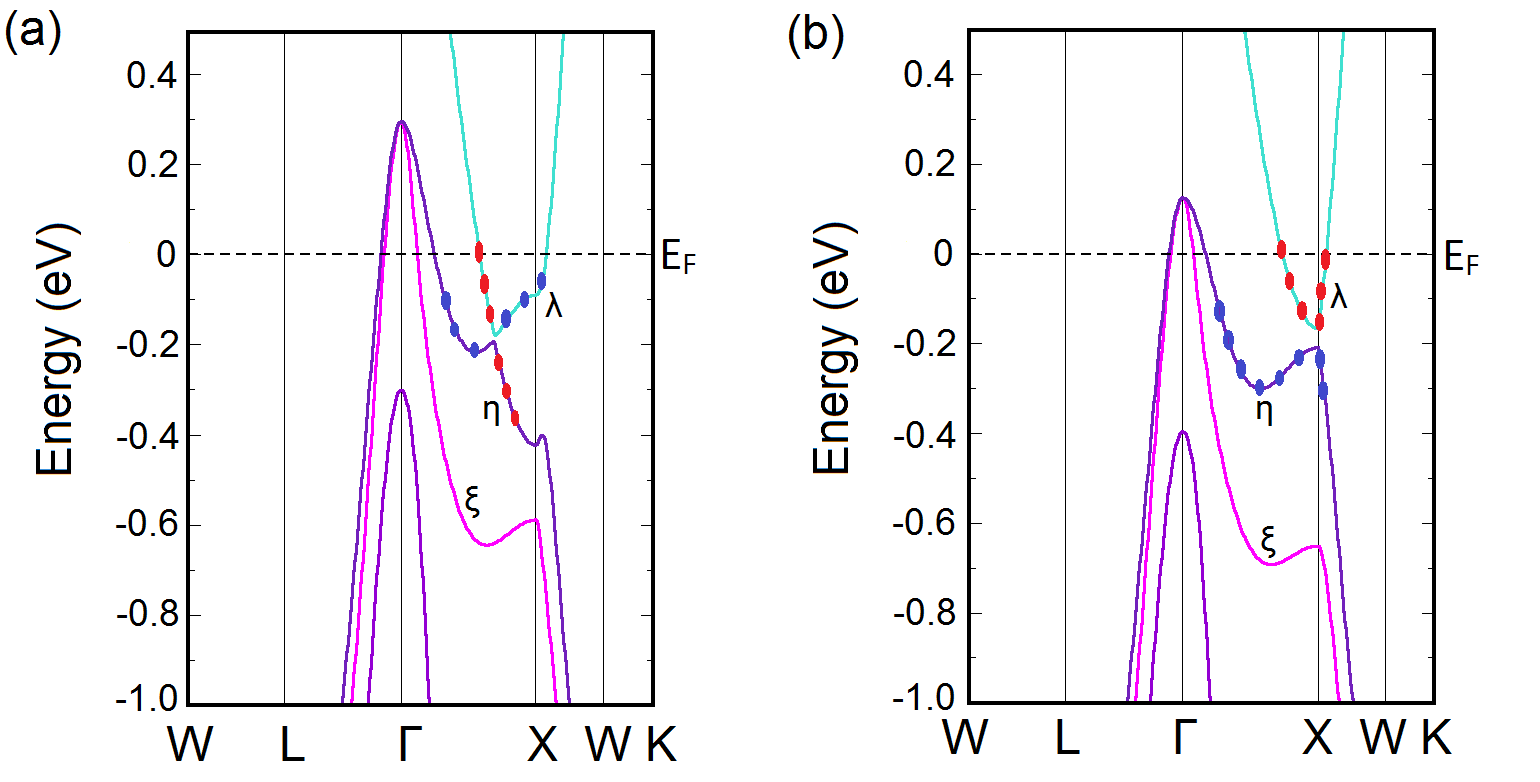} 
\caption{Calculated bulk band structures of LaSb with (a) LDA functional and (b) mBJLDA functional along the high symmetry k-path ($W-L-\Gamma-X-W-K$) shown in Fig.~\ref{LaBi_struct}(b). Here, the red dots indicate the contribution of the La 5d orbitals and the blue dots show the contribution of the Sb 5p orbitals. The band inversion as indicated by the switching of the Sb-5p and La-5d orbitals between the valence and the conduction bands in (a) is removed by the inclusion of the mBJ potential in (b).}
\label{LaSb_bulk}
\end{figure}

From the bulk band structure calculated with LDA functional, as shown in Fig.~\ref{LaSb_bulk}(a), we find that three doubly degenerate bands ($\lambda$, $\eta$ and $\xi$) pass the Fermi level creating one electron pocket ($\lambda$) and two hole pockets ($\eta$ and $\xi$). The conduction band (CB) and the valence band (VB) come very close to each other near the $X$ point when SO is incorporated on the LDA calculation, thus giving rise to a band inversion between the $\Gamma$ and $X$ point. CB and VB change their orbital characters while crossing the point of band inversion. Contribution to the band inversion comes from the parity even La-5d orbitals and the odd parity Sb-5p orbitals as obtained from the band character plots (Fig.~\ref{LaSb_bulk}).\\

However, inclusion of the mBJ potential corrects the gap between the conduction and the valence band locally  and consequently obliterates the band inversion.
Appropriation of the meta-GGA potential results in valence band of p-character and conduction band of d-character with the separation between them
increased, which is in agreement with the previous first-principles calculations of LaSb~\cite{Guo_2016} and supports  the experimental findings of Oinuma
 \textit{et al}.~\cite{Oinuma_2017}
Here, it is important to point out that the LDA as well as GGA approximations tend to underestimate
the band gap. Now, if the material has a small
positive band gap, these approximations may underestimate and result in a negative
value of the band gap, leading to an erroneous prediction of
a  band inversion when in reality there is  none.~\cite{Feng_2010} Therefore, a proper investigation would necessitate a DFT+GW calculation for the proper band separation. However, GW or other hybrid
functionals are computationally very expensive and mBJLDA has been proved to be a much more inexpensive and reliable alternative as shown here.
\subsection*{Surface band structure of LaSb}
\begin{figure}[h]
\centering
\includegraphics[scale=.14]{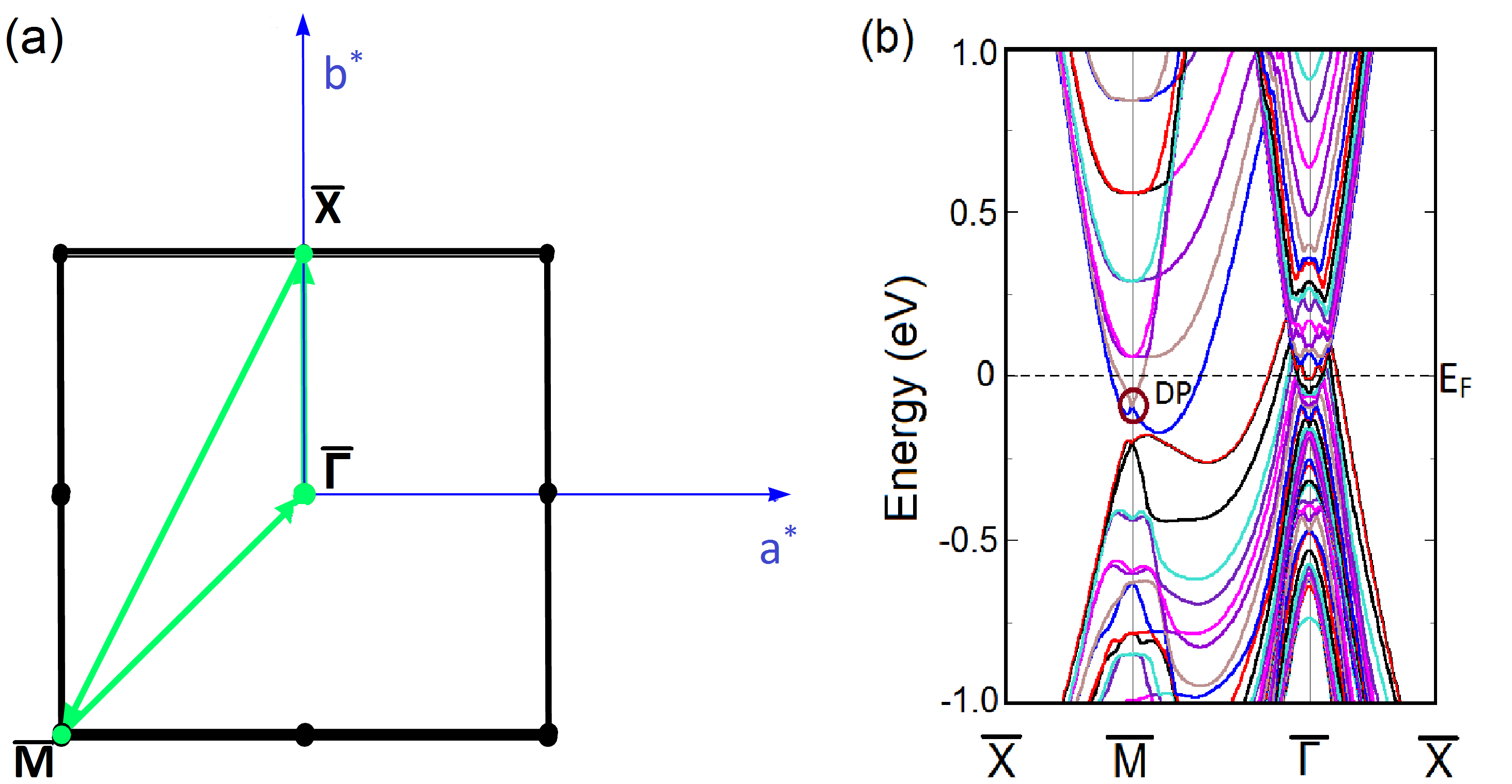}
\caption{(a) Projection of the (001) surface Brillouin zone. $a^{*}$ and $b^{*}$ denote the reciprocal lattice vectors.(b) The surface band structure of LaSb along the high symmetry k-path ($\bar{X}$-$\bar{M}$-$\bar{\Gamma}$-$\bar{X}$) as shown in (a). The Dirac point is encircled and is denoted by DP. Presence of odd number of gapless Dirac cones at the $\bar{M}$-point on the (001) surface suggests that LaSb may have a non-trivial topological character.}
\label{LaSb_surface}
\end{figure}

Although the mBJLDA+SO calculation shows that LaSb is topologically trivial due to the absence of bulk band inversion, controversies still persist regarding the topological nature of LaSb because of the presence of surface Dirac cones previously observed in VUV-ARPES experiments.~\cite{Niu_2016}
This necessitates a thorough study of the LaSb surface states and their response to small perturbations. To that end, we have performed a slab calculation to investigate the (001) surface band dispersion along the $\bar{X}$-$\bar{M}$-$\bar{\Gamma}$-$\bar{X}$ direction using a 18-layer slab separated by 10 \r{A} vacuum. The (001) surface Brillouin zone and band structure are shown in Fig.~\ref{LaSb_surface}.
It is found that although the bulk band structure (using mBJ potential) does not show any band inversion, the (001) projected surface contains a Dirac cone at the $\bar{M}$ point, showing that LaSb may have a non-trivial topological character due to the presence of odd number of Dirac points on the surface.
This result supports the VUV-ARPES experimental findings by Niu \textit{et al}.~\cite{Niu_2016}
\begin{figure}[h]
\centering
\includegraphics[scale=.254]{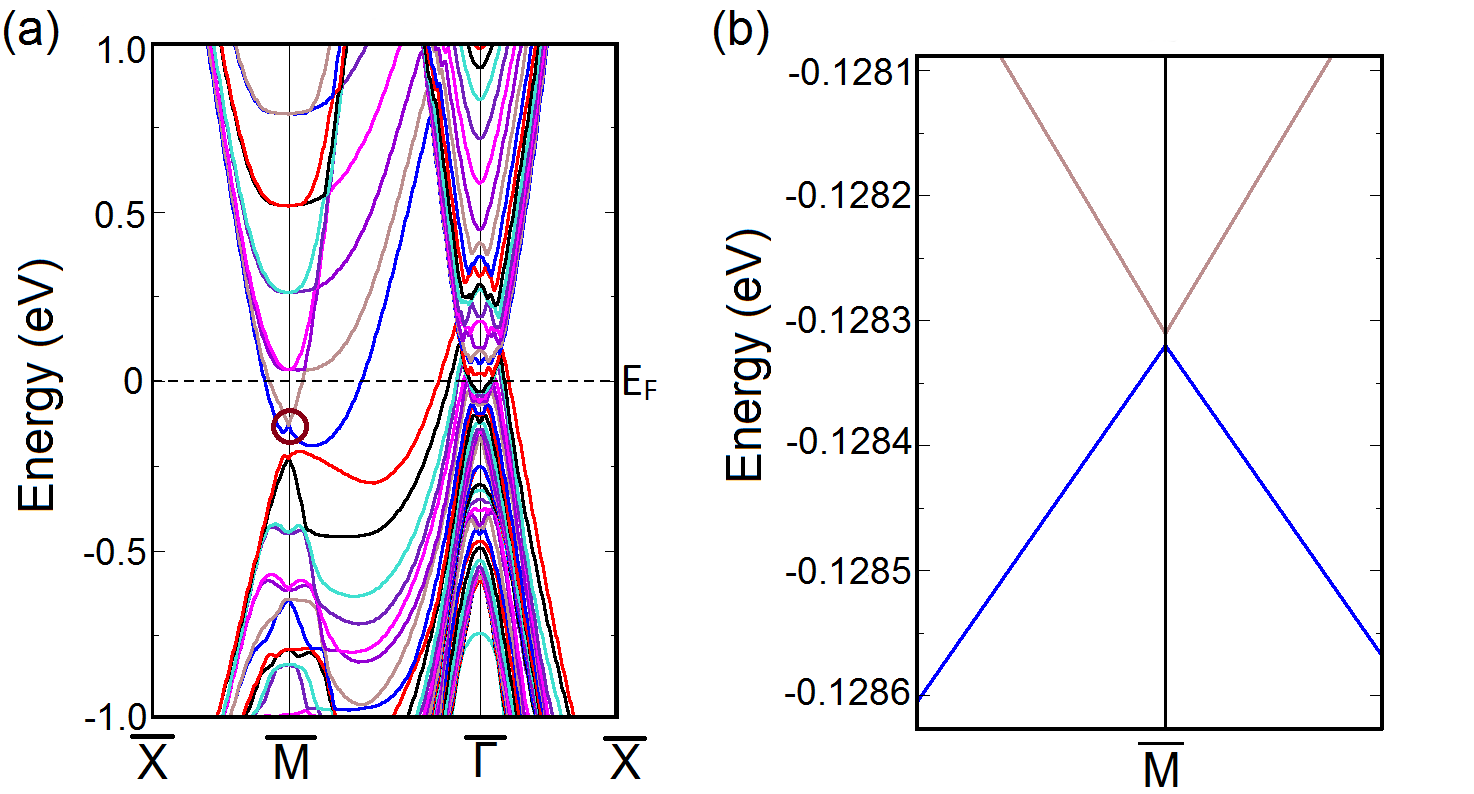} 
\caption{(a) Perturbed (001) surface band structure of LaSb and (b) The corresponding Dirac cone at the $\bar{M}$-point. The Dirac cone becomes gapped on the application of a small TR-preserving perturbation, indicating the Dirac cone at the unperturbed (001) surface (Fig.~\ref{LaSb_surface}(b)) was a result of accidental degeneracy, thus confirming the topologically trivial nature of LaSb.}
\label{LaSb_DC}
\end{figure}

Now for a 3D TR-preserving system to be topologically non-trivial, the most important signature should come from the response of its surface states to TR-preserving perturbations.
 Therefore, if LaSb were topologically non-trivial, the Dirac point found at the (001) surface of LaSb should be robust against small perturbations as long as the perturbation is invariant under time reversal symmetry. In order to check the stability of the Dirac cone at the (001) surface band, we use uniaxial stress as perturbation, which is manifestly invariant under time reversal. We simulate the effect of uniaxial stress in our slab calculations for LaSb by tuning the $c/a$ ratio of the slab.
 Increasing
the $c/a$ ratio by $\sim$2\%, as shown in Fig.~\ref{LaSb_DC}(a), naively it appears that the Dirac cone survives at the $\bar{M}$ point of the surface Brillouin zone of LaSb. However, as shown in Fig.~\ref{LaSb_DC}(b), closer inspection of the Dirac-like gap closing at the $\bar{M}$ point in the surface Brillouin zone
reveals that a non-zero gap ($\sim$ 0.01 meV) appears between the valence and the conduction bands at the $\bar{M}$ point,  which were crossing initially in the absence of the perturbation (see Fig.~\ref{LaSb_surface}).  The accuracy of our slab calculations with and without the perturbation, and reality of the avoided band crossing and associated small gap found between the conduction and valence bands at $\bar{M}$ point in the presence of uniaxial stress, are further confirmed in the next section where we show that similar calculations for LaBi, which is a known $\mathbb{Z}_2$ semimetal,~\cite{Lou_2017,Nayak_2017} preserve both the bulk band inversion and the surface Dirac cone even in the presence of TR-preserving perturbations. 

While commenting on such subtle changes the numerical accuracy of the calculation and the step sizes are of paramount importance. However, since the gap opening takes place at $\bar{M}$-point, a high symmetry point of the slab Brillouin zone, this point is always accounted for. Still we vary the step sizes of the k-mesh and arrive at this unambiguous conclusion.    

\section*{Bulk and surface spectrum of LaBi}
\subsection*{Bulk band structure of LaBi}
\begin{figure}[h]
\centering
\includegraphics[scale=.26]{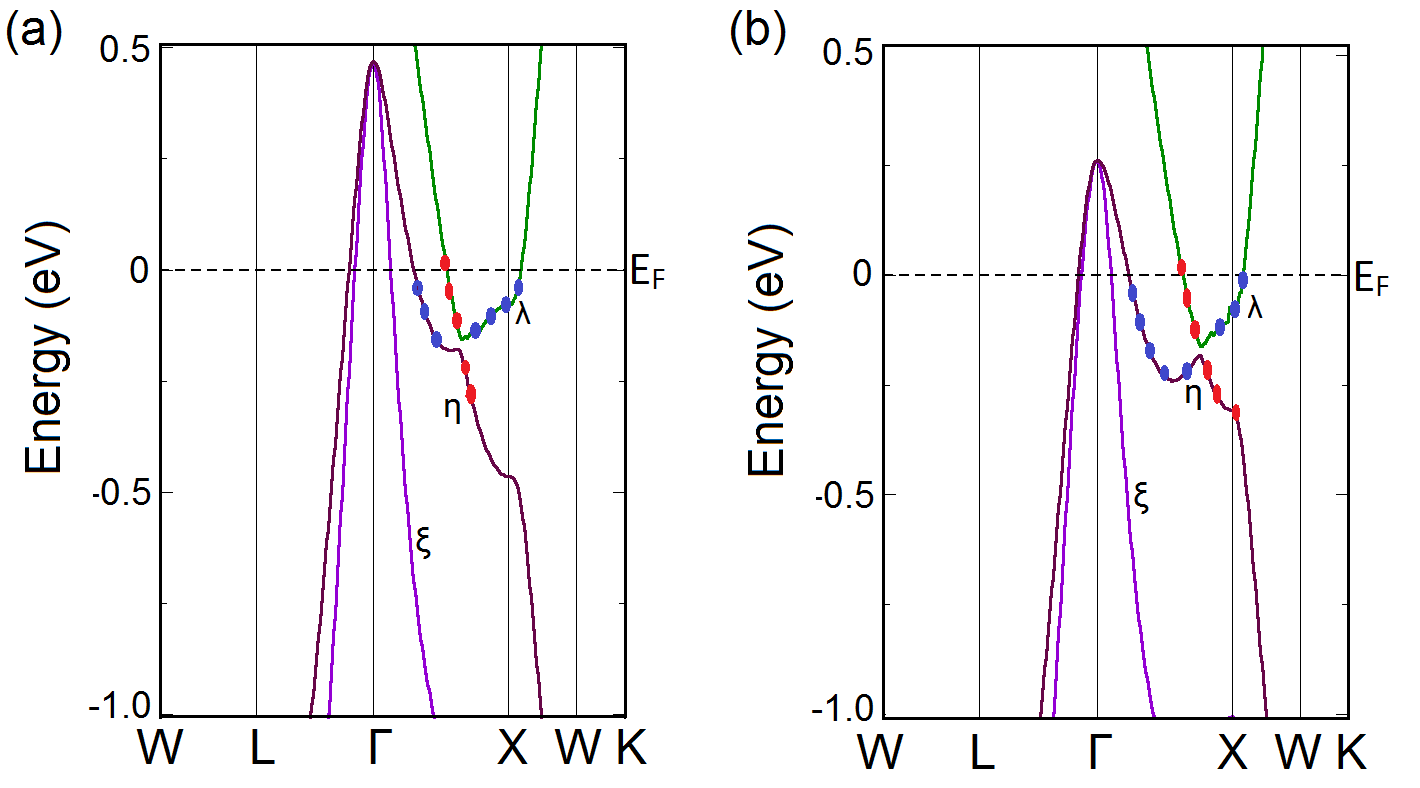}
\caption{Calculated bulk band structures of LaBi with (a) LDA functional and (b) mBJLDA functional along the high symmetry k-path ($W-L-\Gamma-X-W-K$) shown in Fig.~\ref{LaBi_struct}(b). Here, the red dots indicate the contribution of the La 5d orbitals and the blue dots show the contribution of the Bi 6p orbitals. The band inversion as indicated by the switching of the Bi-6p and La-5d orbitals between the valence and the conduction bands in (a) is still preserved and is in fact further consolidated in (b).}
\label{LaBi_bulk}
\end{figure}

To further crosscheck our findings about LaSb, we perform similar calculations on LaBi, which is a known $\mathbb{Z}_{2}$ topological semimetal. LaBi has the same crystal structure as that of LaSb (Fig.~\ref{LaBi_struct}(a)). Our optimized lattice parameter $a$ = 6.493 \r{A} matches well with the experimental value of 6.5799 \r{A}. The band structures calculated along the high symmetry directions of the bulk Brillouin zone (Fig.~\ref{LaBi_struct}(b)) are shown in Fig.~\ref{LaBi_bulk}. Though Bi and Sb belong to the same group (group 15) in the periodic table, Bi has larger atomic radius and is heavier compared to Sb. As a result, the SO coupling strength is larger in LaBi, causing the energy gap between the CB and the VB along the $\Gamma-X$ direction more prominent. 

Similar to the case of LaSb, the inclusion of mBJ potential on the LDA+SO calculation results in modifications in the bulk band structure. As can be seen from Fig.~\ref{LaBi_bulk}, with the mBJ potential the top of the valence band at the $\Gamma$-point is pushed downward and the conduction band minimum near the $X$-point is shifted upward. However, the band inversion is still preserved and is in fact further consolidated by reducing the gap between the conduction band and the valence band as shown in Fig.~\ref{LaBi_bulk}(b).~\cite{Guo_2016} This behavior is very different from the band structure of LaSb, where the inclusion of the mBJ potential removes the bulk band inversion from the Brillouin zone corner as shown in Fig.~\ref{LaSb_bulk}.
\subsection*{Surface band structure of LaBi}
Calculating the (001) surface bands of LaBi along the same $k$-path as shown in Fig.~\ref{LaSb_surface}(a), we find that the surface band structure of LaBi is similar to that of LaSb and it contains a Dirac cone at the $\bar{M}$ point. 
\begin{figure}[h]
\centering
\includegraphics[scale=.25]{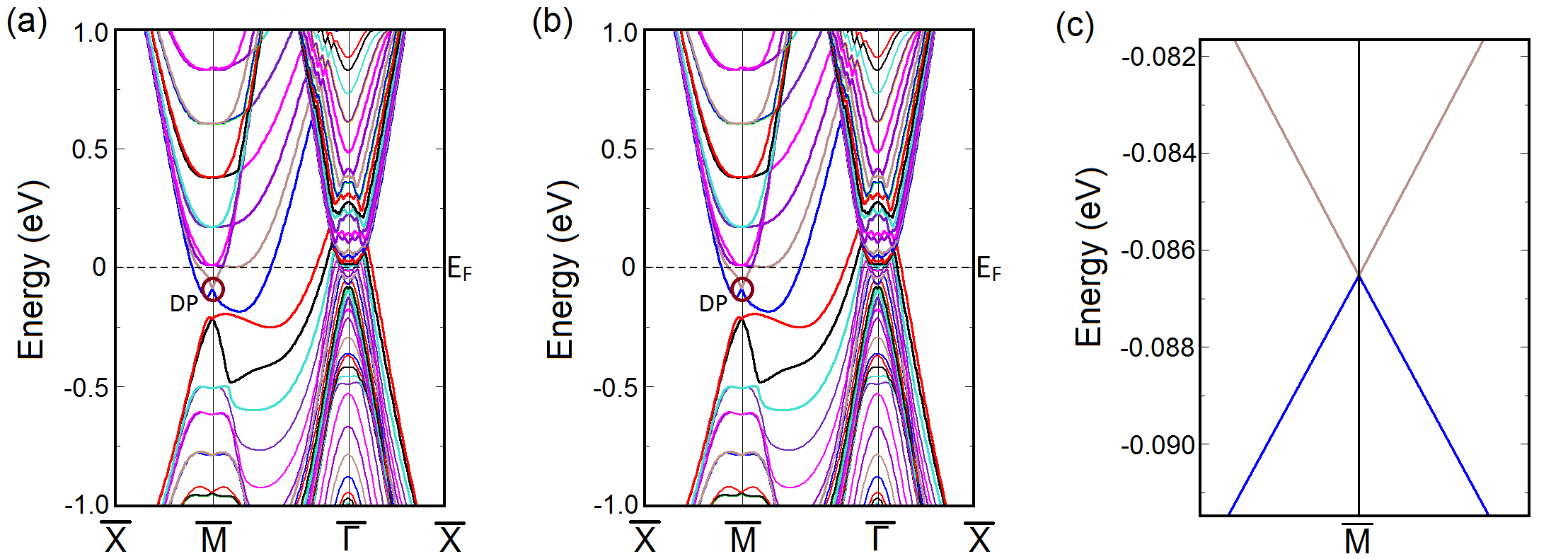} 
\caption{(a) Unperturbed (001) surface band structure of LaBi.  The Dirac point (DP) found at the $\bar{M}$-point is encircled.  (b) (001) surface band structure of LaBi when the $c/a$ ratio of the slab is changed by $\sim$2\%. (c) Corresponding Dirac cone at the $\bar{M}$-point remains stable even after the application of a TR-preserving perturbation. This robustness of the surface states bears testimony to the topological nature of LaBi in contrast to LaSb.}
\label{LaBi_surface}
\end{figure}
In this case, however, we find that applying a uniaxial strain perturbation by way of increasing the $c/a$ ratio in our slab calculations does not significantly change the surface band structure. In particular, we find that the Dirac point at the $\bar{M}$-point in the Brillouin zone remains stable even when subjected to perturbations as long as the system remains time reversal invariant. Since LaBi is topologically non-trivial and possesses a non-zero $\mathbb{Z}_{2}$ index,~\cite{Nayak_2017,Dey_2018} the Dirac point at the surface is robust and remains stable when we change the $c/a$ ratio in our slab calculations (Fig.~\ref{LaBi_surface}(c)). This robustness of the surface states bears testimony to the topological nature of the system. 

By contrast, the Dirac-like surface states of LaSb are fragile and become unstable under small perturbation even in the presence of time reversal symmetry. This proves that the Dirac-cone-like surface states previously observed in both theory and experiment at the (001) surface in LaSb are accidental and LaSb is topologically trivial, i.e., band inversion is absent in bulk LaSb, as also inferred in experiments by Oinuma \textit{et al}.~\cite{Oinuma_2017}

We also perform the (001) surface band structure calculations for two different slabs with 22 layers and 26 layers containing 44 and 52 atoms respectively. Our slab calculations along the high symmetry directions ($\bar{X}$-$\bar{M}$-$\bar{\Gamma}$-$\bar{X}$) for slabs of different thickness show that in all cases the Dirac cone appears at the $\bar{M}$-point (see Supplementary Fig.~S1), thus substantiating that the Dirac cone is a common feature independent of the choice of slab thickness.
\begin{figure}[h]
\centering
\includegraphics[scale=.47]{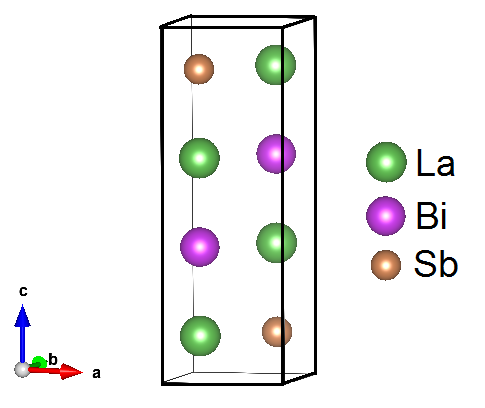}
\caption{1 $\times$ 1 $\times$ 2 supercell of LaSb/LaBi consisting of alternate unit cells of LaSb and LaBi.}
\label{struct}
\end{figure}
\section*{LaSb/LaBi multilayer}
\subsection*{Crystal Structure}
We construct a multilayer by considering a 1 $\times$ 1 $\times$ 2 supercell consisting of alternate unit cells of LaSb and LaBi stacked along the (001) direction as shown in Fig.~\ref{struct}. The multilayer forms a tetragonal structure with space group symmetry $P4/nmm$ (No. 129).
\subsection*{Electronic structure}
\subsubsection*{Bulk energy spectrum of LaSb/LaBi multilayer}
\begin{figure}[h]
\centering
\includegraphics[scale=.13]{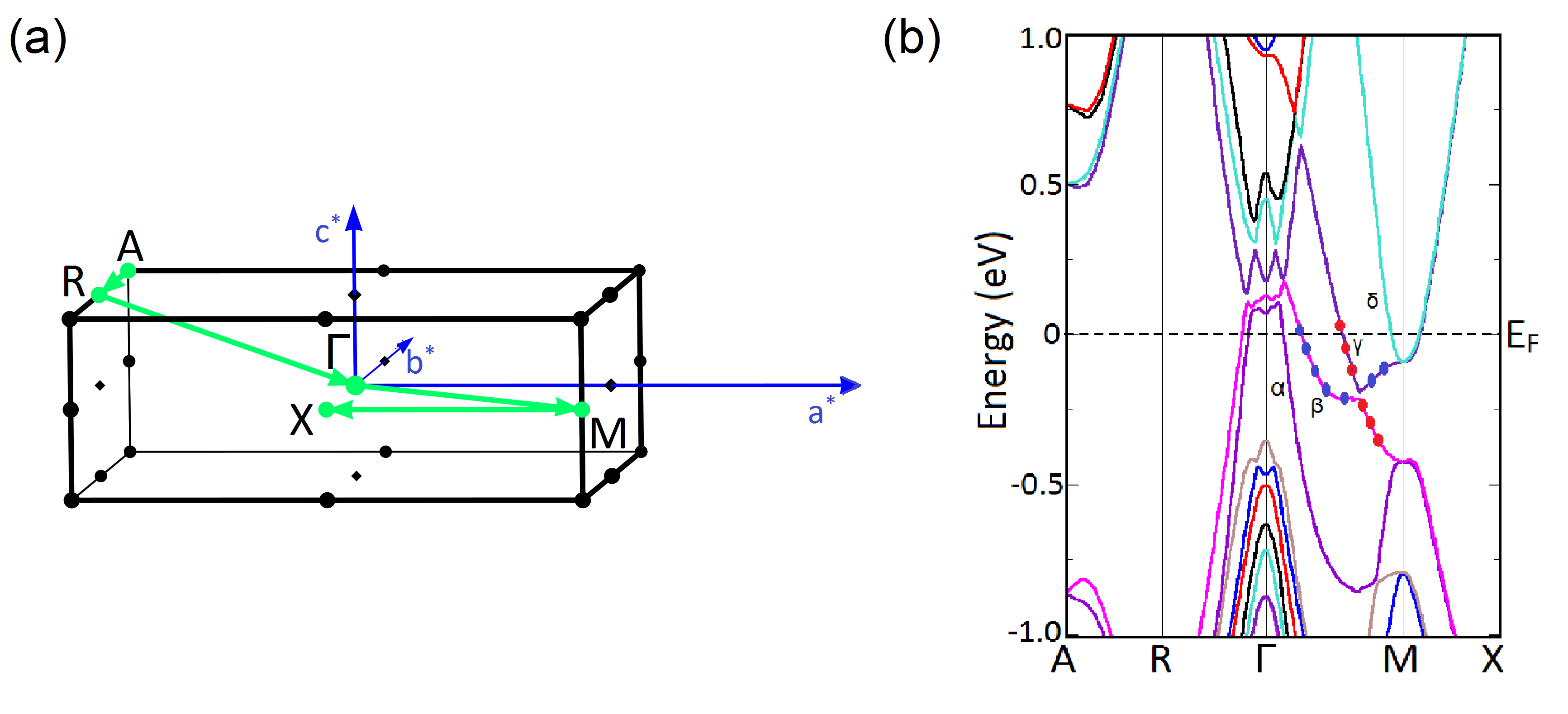} 
\caption{(a) The bulk Brillouin zone of the LaSb/LaBi heterostructure. Here, $a^{*}$, $b^{*}$, $c^{*}$ denote the reciprocal lattice vectors. (b) Bulk band structure along the high symmetry directions ($A-R-\Gamma-M-X$) calculated using mBJLDA functional when the SOC axis is parallel to the (001) direction. The contribution of La-d orbitals are shown by red dots and the contribution of the Bi/Sb-p orbitals are denoted by blue dots. $\beta$ and $\gamma$ band get inverted near the $X$-point as indicated by the switching of the orbital characters along the $\Gamma-M$ direction.}
\label{bulk}
\end{figure}
The bulk band structure of the multilayer is calculated using the optimized lattice parameters $a$ = 4.561 \r{A} and $c$ = 12.899 \r{A} along the high-symmetry directions in the Brillouin zone (Fig.~\ref{bulk}(a)) using the SO incorporated LDA functional. The bulk Brillouin zone and the calculated bulk band structure along the high symmetry directions ($A-R-\Gamma-M-X$) are shown in Fig.~\ref{bulk}. We find that, four doubly degenerate bands cross the Fermi level ($E_F$) creating two hole pockets ($\alpha$ and $\beta$) around the $\Gamma$-point and two electron pockets ($\gamma$ and $\delta$) around the $M$-point, showing the ambipolar nature of the heterostructure. Band character plots of the four bands show that positive parity $\beta$ band and negative parity $\gamma$ band get inverted along the $\Gamma-M$ direction. Contribution to the band inversion comes from La-5d orbitals and the p orbitals of Bi and Sb.
\begin{figure}[h]
\centering
\includegraphics[scale=.25]{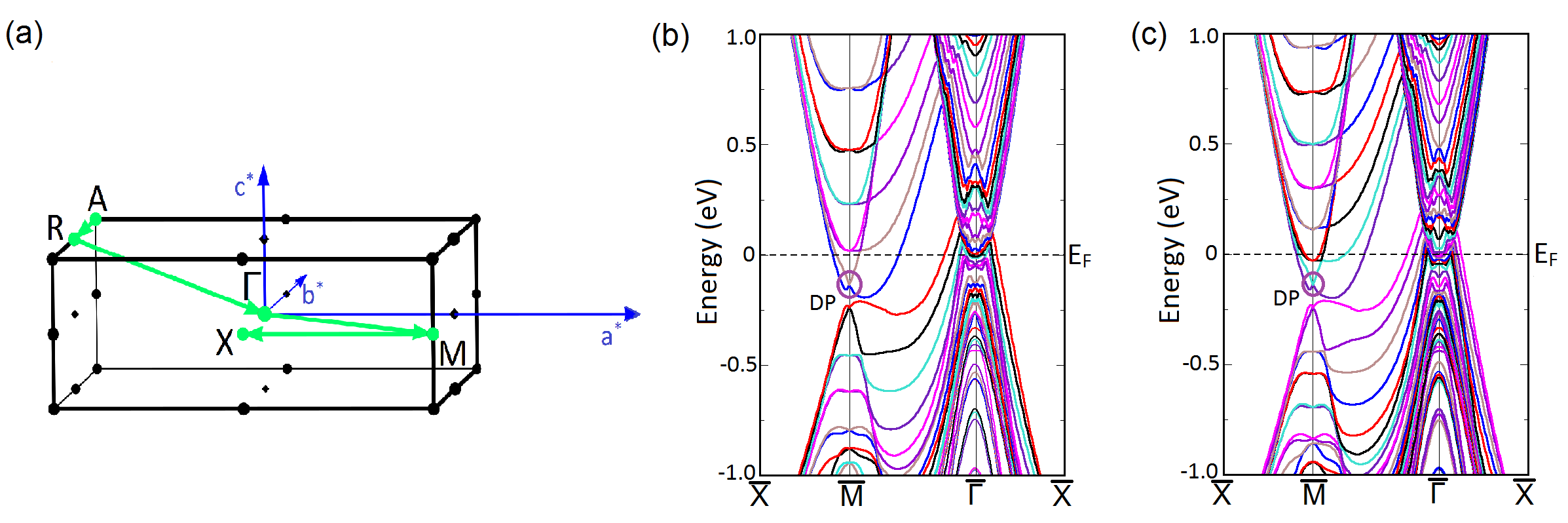}  
\caption{(a) The (001)-projected surface Brillouin zone of the LaSb/LaBi multilayer. $a^{*}$ and $b^{*}$ denote the reciprocal lattice vectors. The surface band dispersions for (b) 16 layer and (c) 20 layer thick slabs along the high symmetry directions ($\bar{X}-{M}-\bar{\Gamma}-X$) shown in (a) calculated using LDA functional and including SO coupling. A Dirac cone appears at the $\bar{M}$-point in each case, which indicates the topologically non-trivial nature of the multilayer. The Dirac point is denoted by DP.}
\label{surface}
\end{figure}

When we include the mBJ functional on our SO incorporated LDA band structure calculations, we find that the valence bands near the $\Gamma$-point shift downwards, while the conduction bands near the $M$-point move upwards. However, the band inversion is still preserved with the inclusion of mBJ potential, indicating the topologically non-trivial nature of the LaSb/LaBi heterostructure.

It has already been theoretically and experimentally verified that LaBi is a $\mathbb{Z}_2$ semimetal with compensated electron and hole contributions~\cite{Guo_2016,Nayak_2017,Lou_2017,Dey_2018}, while LaSb is topologically trivial but exhibits extremely large magnetoresistance~\cite{Guo_2016,Zeng_2016}. Our mBJLDA+SO calculations reveal that their multilayer has a non-trivial topological character. 
\subsubsection*{Surface band structure of LaSb/LaBi multilayer}
Since the mBJLDA+SO calculated bulk band structure shows the presence of band inversion, the probe of surface states becomes indispensable. To confirm the topologically non-trivial character of the LaSb/LaBi heterostructure, we investigate the (001)-projected surface band structure using slab method. We construct 16 layer and 20 layer thick slabs containing 32 atoms and 40 atoms respectively and separated by 20 \r{A} vacuum from the real space. The (001)-projected surface Brillouin zone and the LDA+SO surface band dispersions along the $\bar{X}-{M}-\bar{\Gamma}-X$ direction are shown in Fig.~\ref{surface}. Investigation of Fig.~\ref{surface}(b) and (c) reveals that a massless Dirac cone appears at the $\bar{M}$-point in each case. Appearance of odd number of Dirac cones on the (001) surface suggests that the multilayer consisting of alternate unit cells of topological semimetal LaBi and topologically trivial semimetal LaSb is analogous to the $\mathbb{Z}_2$ semimetal LaBi and possesses a non-trivial topological character.
\begin{figure}[h]
\centering
\includegraphics[scale=.23]{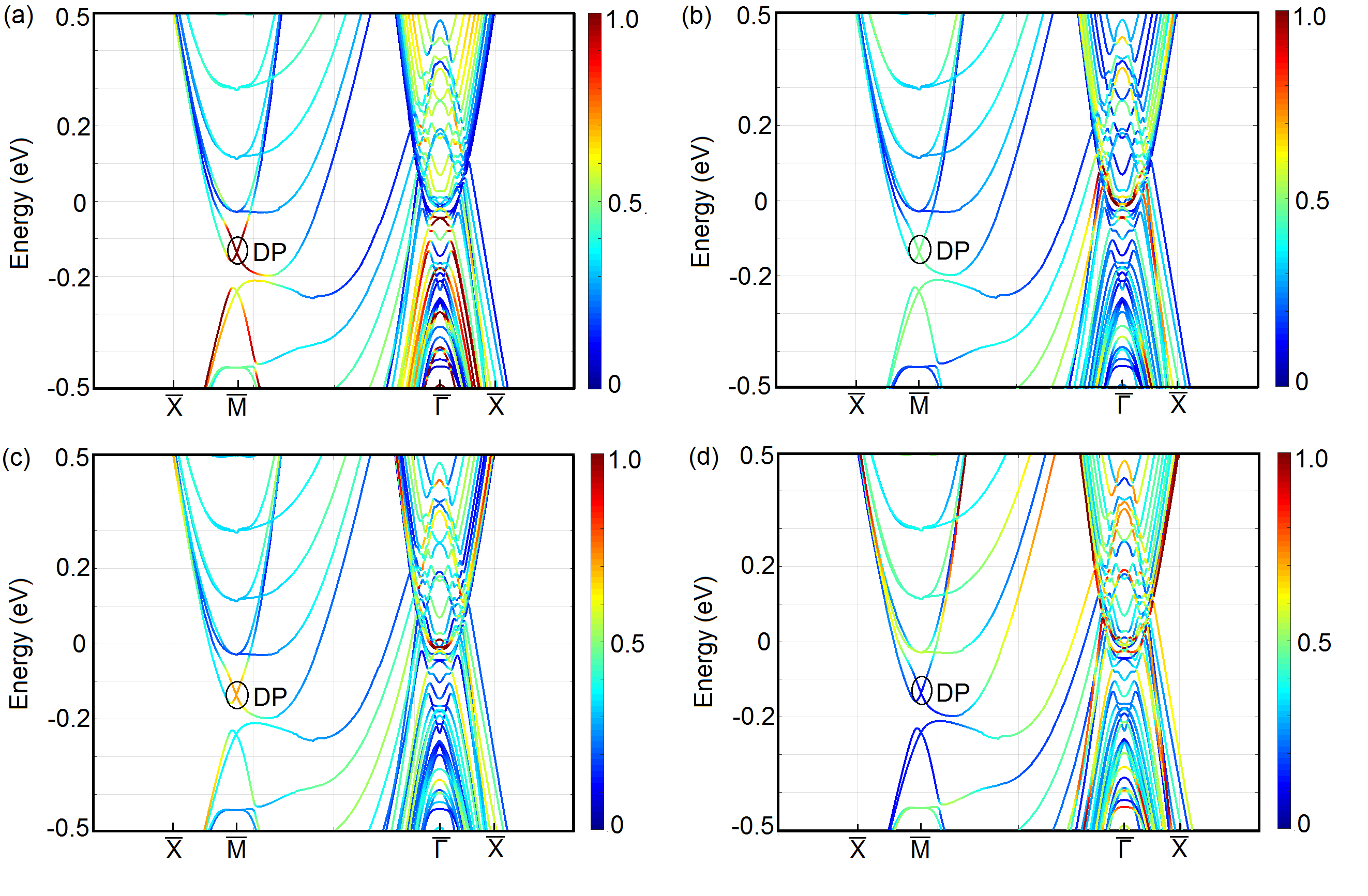}  
\caption{Atomic weights factors of the bands shown in Fig.~\ref{surface}(c) for a 20 layer thick slab. The atomic weights for the (a) topmost, (b) second topmost, (c) third topmost and the (d) central bulk layer are given by the line intensity. As seen, the intensity at the Dirac point (DP) is maximum for the topmost (surface) layer i.e. the contribution to the Dirac cone bands at the $\bar{M}$ point mainly comes from the atoms in the topmost (surface) layer, indicating the surface-state origin of the Dirac cone observed at the (001) surface of the LaSb/LaBi multilayer.}
\label{atomic_weight}
\end{figure}

In order to verify the surface-state origin of the Dirac-cone-like feature observed at the (001) surface, we evaluate the atomic weight factors of the bands shown in Fig.~\ref{surface}(c). Calculations of the atomic weight factors for the topmost, second topmost, third topmost and the central bulk layer reveal that the contribution to the Dirac cone bands at the $\bar{M}$ point mainly comes from the atoms in the topmost (surface) layer. 
In Fig.~\ref{atomic_weight}, the atomic weights of the bands are shown by the line intensity. As seen, the intensity at the Dirac point (DP) is maximum for the topmost (surface) layer, indicating the surface-state origin of the Dirac cone, which in turn implies that the Dirac cone observed at the (001) surface of the LaSb/LaBi multilayer is topologically protected and will be robust against time-reversal-preserving perturbations.

Applying a small time-reversal-preserving perturbation to the surface states of the multilayer in the form of a uniaxial strain, we find that the Dirac cone at the (001) surface $\bar{M}$-point remains unaffected. The robustness of the surface states against small TR-preserving perturbations ($\sim$ 2\% change in the c/a ratio of the slab) confirms the topologically non-trivial nature of the multilayer containing alternate unit cells of LaSb and LaBi.

\subsection*{Calculation of $\mathbb{Z}_2$ index}
In 3D, there are four $\mathbb{Z}_2$ indices that characterize a topological insulator and distinguish two different types of topological classes, strong topological insulators (STI) and weak topological insulators (WTI).~\cite{Fu:Inversion} The topological properties of the STIs are robust against time-reversal-preserving perturbations, whereas, the topological character of the WTIs can be destroyed by disorders. According to Fu and Kane,~\cite{Fu:Inversion} $\mathbb{Z}_2$ index of a 3D topological insulator can be calculated by considering the parity product of the valence bands at the eight time-reversal-invariant momentum (TRIM) points if the system preserves both time-reversal symmetry (TRS) and inversion symmetry (IS).

Investigation of the bulk band structures of LaSb, LaBi and LaSb-LaBi heterostructure (Fig.~\ref{LaSb_bulk}, Fig.~\ref{LaBi_bulk} and Fig.~\ref{bulk}) reveals that at each $k$-point the conduction band and the valence band are gapped, though there is no indirect gap at the Fermi level. This allows us to define $\mathbb{Z}_2$ index for these materials. Therefore, to verify the topological character of LaSb, LaBi and their multilayer, we calculate the first $\mathbb{Z}_2$ topological invariant $\nu_0$, which is robust against time-reversal-preserving perturbations, using the following relation~\cite{Fu:Inversion} :
\begin{equation}
{(-1)}^{\nu_0} = \prod_{n=1}^{8} \delta_n
\label{Z2}
\end{equation}
where, $\delta_n$ is the parity product of the valence bands at the $n$-th TRIM point.
\begin{table}[h]
  \centering
  \caption{Calculation of $\mathbb{Z}_2$ index for LaSb, LaBi and their heterostructure with LDA+SO and mBJLDA+SO functionals using Eq.~\ref{Z2}. As seen, LaBi~\cite{Nayak_2017,Dey_2018} and LaSb/LaBi heterostructure possess a non-trivial $\mathbb{Z}_2$ index $\nu_0$ = 1, indicating their non-trivial topological character. On the other hand, $\nu_0$ changes from 1 to 0 for LaSb, when we introduce the mBJ potential, thus confirming the topologically trivial nature of LaSb.}
  \label{tab1}
  \begin{tabular}{|c|c c c| c c c| c c c|}
    \hline
    &\multicolumn{3}{|c|}{LaSb}& \multicolumn{3}{c|}{LaBi} & \multicolumn{3}{|c|}{LaSb/LaBi multilayer}\\
    \hline
     &TRIM points & $\delta_m$ & $\nu_0$ & TRIM points & $\delta_m$ & $\nu_0$ & TRIM points & $\delta_m$ & $\nu_0$ \\
    \hline
    & &  & &  &   & & 1$\Gamma$ & $-$ &\\
    & &  &  &  &   &  & 1$A$ & $+$ &\\
    with LDA+SO&1$\Gamma$ & $+$&  & 1$\Gamma$ & $+$  &  & 1$M$ & $-$ &\\
    functional& 4$L$& $+$ & 1 & 4$L$  & $+$ & 1 & 2$R$ & $+$ &1\\
    & 3$X$& $-$ &   & 3$X$  & $-$ &  & 2$X$ & $+$ &\\
    & &  &  &  &  &  & 1$Z$ & $-$ &\\
    \hline
    & &  & &  &   & & 1$\Gamma$ & $-$ &\\
    & &  &  &  &   &  & 1$A$ & $+$ &\\
    with mBJLDA+SO&1$\Gamma$ & $+$&  & 1$\Gamma$ & $+$  &  & 1$M$ & $-$ &\\
    functional& 4$L$& $+$ & 0 & 4$L$  & $+$ & 1 & 2$R$ & $+$ &1\\
    & 3$X$& $+$ &   & 3$X$  & $-$ &  & 2$X$ & $+$ &\\
    & &  &  &  &  &  & 1$Z$ & $-$ &\\
    \hline
\end{tabular}
\end{table}

For the calculation of $\nu_0$, we consider only the $\lambda$, $\eta$ and $\xi$ bands of LaSb and LaBi (in Fig.~\ref{LaSb_bulk} and Fig.~\ref{LaBi_bulk}) and the $\alpha$, $\beta$, $\gamma$, $\delta$ bands of the LaSb/LaBi heterostructure (as given in Fig.~\ref{bulk}(b)), since other bands are isolated and are topologically trivial. FCC Brillouin zone (Fig.~\ref{LaBi_struct}(b)) contains three sets of TRIM points : one $\Gamma$, four $L$ and three $X$. On the other hand, the TRIM points of a tetragonal Brillouin zone (Fig.~\ref{bulk}(a)) include one $\Gamma$, one $A$, one $M$, two $R$, two $X$ and one $Z$. Calculating the parity products ($\delta_n$) of the relevant bands at the TRIM points, we determine the $\mathbb{Z}_2$ invariant for LaSb, LaBi and LaSb-LaBi heterostructure, as shown in Table~\ref{tab1}. From Table~\ref{tab1}, we find that LaBi~\cite{Nayak_2017,Dey_2018} and LaSb/LaBi multilayer possess a non-trivial $\mathbb{Z}_2$ topological invariant when we use both LDA and mBJLDA functionals, which is consistent with our (001) surface band structure calculations. However, in case of LaSb, inclusion of the mBJLDA functional changes the $\mathbb{Z}_2$ index from 1 to 0, confirming that LaBi and LaSb-LaBi multilayer are topologically non-trivial, whereas, LaSb is topologically trivial with accidental Dirac cones on the (001) surface. The parities of the relevant bands are shown in Supplementary Fig.~S2.
\section*{Summary and Conclusion}
In conclusion, we have examined the presence or absence of an odd number of band inversions between bands of different parities in the bulk band structure of LaSb, LaBi and their multilayer, along with the stability of unusual Dirac-cone-like surface states against time reversal symmetric perturbations. We have determined the $\mathbb{Z}_2$ invariant for all our systems to validate our conclusions from the bulk and surface band structure calculations. The presence of an odd number of band inversions between bands of different parities in parity-symmetric systems~\cite{Fu:Inversion} mathematically translates into a non-trivial $\mathbb{Z}_2$ index. A non-trivial $\mathbb{Z}_2$ index in turn implies the presence of topologically protected gapless surface states at boundaries separating the bulk spectrum from vacuum, which can be viewed as a topologically trivial insulator. Conversely, the absence of band inversion in the bulk insulator should imply topologically trivial surface states, if at all, which should be extremely sensitive to perturbations even in the presence of time reversal symmetry. We apply these ideas to resolve the existing disagreements among various groups regarding the topological property of LaSb comparing it with LaBi, which is a known $\mathbb{Z}_2$ semimetal, by first-principles calculations.

 We have calculated the bulk band structure of LaSb using both LDA and mBJLDA exchange functionals including the effect of spin-orbit coupling. We find that the band inversion is wiped out in LaSb when we include the mBJ potential, which was initially present when we used the LDA functional. The absence of band inversion in the bulk suggests that LaSb should not have topologically protected surface states. However, the (001) projected surface band structure calculations of LaSb show the presence of odd number of Dirac cones, a result also supported experimentally.~\cite{Niu_2016} More recently, bulk sensitive ARPES experiments using
soft-x-ray photons by Oinuma \textit{et al}~\cite{Oinuma_2017} have shown that there is no bulk band inversion present in LaSb and CeSb, contradicting the previous VUV-ARPES experiments~\cite{Niu_2016} which found unusual Dirac-cone-like surface states in LaSb. Motivated by these existing disagreements, we revisit the question of topological band structure of LaSb and LaBi, focusing in particular on the bulk band inversion and the  stability of surface Dirac cones.  We find that not only does the bulk band inversion disappears in LaSb upon inclusion of the mBJ potential, but also a time-reversal-preserving perturbation such as uniaxial strain in the form of a change in the $c/a$ ratio in our slab calculation removes the linearly
dispersive Dirac cone from the (001) surface when the ratio is increased by $\sim$2\%. However, similar calculations for LaBi show that the band inversion in the bulk band structure is unaffected by the introduction of the mBJ potential. Furthermore, we find that a time-reversal-symmetric perturbation such as  uniaxial strain on the (001) surface states of LaBi does not change the shape or nature of the Dirac cone located at the $\bar{M}$ point. These calculations are consistent with the topologically non-trivial nature of LaBi, which is a 3D topological $\mathbb{Z}_2$ semimetal, while we conclude that the extreme magnetoresistive compound LaSb is topologically trivial, with unusual Dirac-cone-like surface states which can only be accidental.

On the other hand, our mBJLDA+SO calculations show that a multilayer formed by alternating unit cells of LaSb and LaBi has clear band inversion near the $M$-point in the bulk band structure. 
Investigation of the (001) surface states of the heterostructure confirms the presence of a Dirac cone at the (001) $\bar{M}$-point in the surface Brillouin zone of the multilayer. We also confirm the robustness of the surface Dirac cone to TR-preserving perturbations, establishing the topologically non-trivial properties of the multilayer, which can potentially be used for device applications by chemical substitution and/or alteration of the stacking sequence of multilayer.
This kind of topologically protected heterostructure, with possible extreme magnetoresistive properties co-existing with topologically non-trivial band structure, provides us with an avenue to design new materials with novel functionalities that can be engineered in the laboratory. 

\section*{Methods}
In order to determine the presence or absence of bulk band inversion in LaSb, LaBi and their heterostructure, we adopt the all electron full potential linearized augmented plane wave (FLAPW) method for performing the band structure calculations using WIEN2K
code.~\cite{Blaha_2002} For the exchange correlation part, local density approximation (LDA) and modified Becke Johnson (mBJ)
potential~\cite{Becke_2006} are used. At first, we have optimized the structures using LDA functional to get the equilibrium lattice parameters.
The mBJ corrects the conventional
LDA or GGA type of exchange correlations by
incorporating the effect of hole or unoccupied states, hence  improving
the separation between the levels present
near the Fermi level (E$_\text{F}$). mBJLDA has proved to be a much cheaper alternative to very expensive GW calculations,
while yielding  almost similar accuracy.~\cite{Tran_2009} To that end, we have performed mBJLDA calculations on the optimized structures. 

Effect of spin-orbit (SO) coupling has been accounted for through a second
variational procedure, where
states up to 9 Rydberg (Ry) above E$_\text{F}$ are included in the basis
expansion, and the relativistic p$_{1/2}$ corrections have  been incorporated
for the 5p and 6p orbitals to enhance the
accuracy. A $ 20 \times 20 \times 20 $ k-mesh is used for the whole Brillouin zone. For the partial waves inside the muffin tin spheres we use $L_{max} = 12$ and for the charge Fourier expansion we take $G_{max} = 14$ ${\text{bohr}}^{-1}$. For higher accuracy, the radius of the muffin tin ($RMT$) spheres are chosen such that $ K_{max} \times RMT$ = 9.5, where $K_{max}$ is the plane wave momentum cut-off.
We have done fat band analysis to obtain the orbital character of the bands. 

Finally, to calculate the surface band dispersions of LaSb, LaBi and the LaSb-LaBi multilayer, we have performed slab calculations using a thick slab separated by vacuum from the supercells in the real space.

\section*{Acknowledgments}
UD and AT appreciate access to the computing facilities of the DST-FIST (phase-II) project installed in the Department of Physics, IIT Kharagpur, India.
ST acknowledges support from ARO Grant No: (W911NF-16-1-0182). UD would like to acknowledge the Ministry of Human Resource Development (MHRD) for research fellowship.
\section*{Author contributions statement}
ST conceived the problem, UD performed all the calculations, all authors analysed the results, UD, MC, AT and ST wrote the paper. All authors reviewed the manuscript. 
\section*{Additional information}
There are no competing interests associated with this paper.


\begin{thebibliography}{10}
\bibitem{Mele} Kane, C. L. \& Mele, E. J. $Z_2$ topological order and the quantum spin hall effect. \textit{Phys. Rev. Lett.} \textbf{95,} 146802 (2005).
\bibitem{Moore} Moore, J. E. \& Balents, L. Topological invariants of time-reversal-invariant band structures. \textit{Phys. Rev. B} \textbf{75,} 121306(R)
(2007).
\bibitem{Fu:2007} Fu, L., Kane, C. L., \& Mele, E. J. Topological insulators in three dimensions. \textit{Phys. Rev. Lett.} \textbf{98,} 106803 (2007).
\bibitem{Roy} Roy, R. $Z_2$ classification of quantum spin hall systems: an approach using time-reversal invariance. Phys. Rev. B 79, 195321
(2009).
\bibitem{Bernevig} Bernevig, B. A., Hughes, T. L. \& Zhang, S. C. Quantum spin hall effect and topological phase transition in HgTe quantum
wells. \textit{Science} \textbf{314}, 1757 (2006).
\bibitem{Konig} K\"{o}nig, M. \textit{et al.} Quantum spin hall insulator state in HgTe quantum wells. \textit{Science} \textbf{318,} 766 (2007).
\bibitem{Hsieh} Hsieh, D. \textit{et al.} A topological dirac insulator in a quantum spin hall phase. \textit{Nature} \textbf{452,} 970 (2008).

\bibitem{Kane_2010} Hasan, M. Z. \& Kane, C. L. Colloquium: topological insulators. \textit{Rev. Mod. Phys.} \textbf{82,} 3045 (2010).

\bibitem{Zhang_2011} Qi, X. L. \& Zhang, S. C. Topological insulators and superconductors. \textit{Rev. Mod. Phys.} \textbf{83,} 1057 (2011).


\bibitem{Peng_2010} Peng, H. L. \textit{et al.} Aharonov-bohm interference in topological insulator nanoribbons. \textit{Nat. Mater.} \textbf{9,} 225 (2010).

\bibitem{Chen_2010} Chen, J. \textit{et al.} Gate-voltage control of chemical potential and weak antilocalization in Bi$_2$Se$_3$. \textit{Phys. Rev. Lett.} \textbf{105,} 176602 (2010).

\bibitem{He_2011} He, H. T. \textit{et al.} Impurity effect on weak antilocalization in the topological insulator Bi$_2$Te$_3$. \textit{Phys. Rev. Lett.} \textbf{106,} 166805 (2011).

\bibitem{Hor_2011} Checkelsky, J. G., Hor, Y. S., Cava, R. J. \& Ong, N. P. Bulk band gap and surface state conduction observed in voltage-tuned
crystals of the topological insulator Bi$_2$Se$_3$. \textit{Phys. Rev. Lett.} \textbf{106,} 196801 (2011).

\bibitem{Qu_2010} Qu, D. X., Hor, Y. S., Xiong, J., Cava, R. J. \& Ong, N. P. Quantum oscillations and hall anomaly of surface states in the
topological insulator Bi$_2$Te$_3$. \textit{Science} \textbf{329,} 821 (2010).









\bibitem{Ando_2013} Ando, Y. Topological insulator materials. \textit{J. Phys. Soc. Jpn.} \textbf{82,} 102001 (2013).

\bibitem{Bansil_2012} Hsieh, T. H. \textit{et al.} Topological crystalline insulators in the SnTe material class. \textit{Nat. Commun.} \textbf{3,} 982 (2012).

\bibitem{Tanaka_2012} Tanaka, Y. \textit{et al.} Experimental realization of a topological crystalline insulator in SnTe. \textit{Nat. Phys.} \textbf{8,} 800 (2012).

\bibitem{Hasan_2012} Xu, S.-Y. \textit{et al.} Observation of a topological crystalline insulator phase and topological phase transition in Pb$_{(1-x)}$Sn$_{(x)}$Te. \textit{Nat. Commun.} \textbf{3,} 1192 (2012).

\bibitem{Story_2012} Dziawa, P. \textit{et al.} Topological crystalline insulator states in Pb$_{1-x}$Sn$_x$Se. \textit{Nat. Mater.} \textbf{11,} 1023 (2012).

\bibitem{Hasan_2014} Neupane, M. \textit{et al.} Observation of a three-dimensional topological dirac semimetal phase in high-mobility Cd$_3$As$_2$. \textit{Nat.
Commun.} \textbf{5,} 3786 (2014).
\bibitem{Gibson_2014} Borisenko, S. \textit{et al.} Experimental realization of a three-dimensional dirac semimetal. \textit{Phys. Rev. Lett.} \textbf{113,} 027603 (2014).

\bibitem{Liu_2014} Liu, Z. K. \textit{et al.} Discovery of a three-dimensional topological dirac semimetal, Na$_3$Bi. \textit{Science} \textbf{343,} 864 (2014).

\bibitem{Hasan_2015} Xu, S.-Y. \textit{et al.} Discovery of a weyl fermion semimetal and topological fermi arcs. \textit{Science} \textbf{349,} 613 (2015).
\bibitem{Ma_2015} Lv, B. Q. \textit{et al.} Experimental discovery of weyl semimetal TaAs. \textit{Phys. Rev. X} \textbf{5,} 031013 (2015).

\bibitem{Yang_2015} Yang, L. X. \textit{et al.} Weyl semimetal phase in the non-centrosymmetric compound TaAs. \textit{Nat. Phys.} \textbf{11,} 728 (2015).
\bibitem{Ando_2016} Souma, S. \textit{et al.} Direct observation of nonequivalent fermi-arc states of opposite surfaces in the noncentrosymmetric weyl
semimetal NbP. \textit{Phys. Rev. B} \textbf{93,} 161112(R) (2016).
\bibitem{Schoop_2016} Schoop, L. M. \textit{et al.} Dirac cone protected by non-symmorphic symmetry and three-dimensional dirac line node in ZrSiS. \textit{Nat. Commun.} \textbf{7,} 11696 (2016).

\bibitem{Takane_2016} Takane, D. \textit{et al.} Dirac-node arc in the topological line-node semimetal HfSiS. \textit{Phys. Rev. B} \textbf{94,} 121108(R) (2016).
\bibitem{Fu:Inversion} Fu, L. \& Kane, C. L. Topological insulators with inversion symmetry. \textit{Phys. Rev. B} \textbf{76,} 045302 (2007).

\bibitem{Bansil_2015} Zeng, M. \textit{et al.} Topological semimetals and topological insulators in rare earth monopnictides. \textit{arXiv e-prints} 1504.03492
(2015).

\bibitem{Guo_2016} Guo, P.-J., Yang, H.-C., Zhang, B.-J., Liu, K. \& Lu, Z.-Y. Charge compensation in extremely large magnetoresistance
materials LaSb and LaBi revealed by first-principles calculations. \textit{Phys. Rev. B} \textbf{93,} 235142 (2016).

\bibitem{Guo_2017} Guo, C. \textit{et al.} Possible weyl fermions in the magnetic kondo system CeSb. \textit{npj Quantum Materials} \textbf{2,} 39 (2017).

\bibitem{Neupane_2016} Neupane, M. \textit{et al.} Observation of dirac-like semi-metallic phase in NdSb. \textit{J. Phys. Condens. Matter} \textbf{28,} 23
(2016).
\bibitem{Felser_2016} Kumar, N. \textit{et al.} Observation of pseudo-two-dimensional electron transport in the rock salt-type topological semimetal
LaBi. \textit{Phys. Rev. B} \textbf{93,} 241106 (2016).

\bibitem{Kaminski_2017} Wu, Y. \textit{et al.} Electronic structure of RSb (R=Y, Ce, Gd, Dy, Ho, Tm, Lu) studied by angle-resolved photoemission
spectroscopy. \textit{Phys. Rev. B} \textbf{96,} 035134 (2017).

\bibitem{Gibson_2015} Tafti, F., Gibson, Q., Kushwaha, S., Haldolaarachchige, N. \& Cava, R. Resistivity plateau and extreme magnetoresistance
in LaSb. \textit{Nat. Phy.} \textbf{12,} 272–277 (2016).

\bibitem{Sun_2016} Sun, S., Wang, Q., Guo, P.-J., Liu, K. \& Lei, H. Large magnetoresistance in LaBi: origin of field-induced resistivity upturn
and plateau in compensated semimetals. \textit{New Journal of Physics} \textbf{18,} 082002 (2016).

\bibitem{Nayak_2017} Nayak, J. \textit{et al.} Multiple Dirac cones at the surface of the topological metal LaBi. \textit{Nat. Commun.} \textbf{8,} 13942 (2017).
\bibitem{Lou_2017} Lou, R. \textit{et al.} Evidence of topological insulator state in the semimetal LaBi. \textit{Phys. Rev. B} \textbf{95,} 115140 (2017).
\bibitem{Niu_2016} Niu, X. \textit{et al.} Presence of exotic electronic surface states in LaBi and LaSb. \textit{Phys. Rev. B} \textbf{94,} 165163 (2016).

\bibitem{Zeng_2016} Zeng, L.-K. \textit{et al.} Compensated semimetal LaSb with unsaturated magnetoresistance. \textit{Phys. Rev. Lett.} \textbf{117,} 127204 (2016).

\bibitem{Oinuma_2017} Oinuma, H. \textit{et al.} Three-dimensional band structure of LaSb and CeSb: absence of band inversion. \textit{Phys. Rev. B} \textbf{96,} 041120 (2017).

\bibitem{Feng_2010} Feng, W., Xiao, D., Zhang, Y. \& Yao, Y. Half-heusler topological insulators: a first-principles study with the tran-blaha
modified becke-johnson density functional. \textit{Phys. Rev. B} \textbf{82,} 235121 (2010).

\bibitem{Dey_2018} Dey, U. Comparative study of the compensated semi-metals LaBi and LuBi: a first-principles approach. \textit{J. Phys. Condens. Matter} \textbf{30,} 205501 (2018).

\bibitem{Blaha_2002} Schwarz, K., Blaha, P. \& Madsen, G. Electronic structure calculations of solids using the WIEN2k package for material
sciences. \textit{Computer Physics Communications} \textbf{147,} 71 (2002).

\bibitem{Becke_2006} Becke, A. D. \& Johnson, E. R. A simple effective potential for exchange. \textit{The Journal of Chemical Physics} \textbf{124,} 221101
(2006).

\bibitem{Tran_2009} Tran, F. \& Blaha, P. Accurate band gaps of semiconductors and insulators with a semilocal exchange-correlation potential.
\textit{Phys. Rev. Lett.} \textbf{102,} 226401 (2009).


\end{thebibliography}
\end{document}